\documentclass[preprintnumbers,twocolumn,secnumarabic,amssymb, nobibnotes, aps, prd]{revtex4}

\usepackage{color}
\usepackage{graphicx}
\usepackage{amsmath}

\usepackage{amssymb}

\begin{document}
\title{Scattering and momentum space entanglement}

\author{Gianluca Grignani}
\affiliation{Department of Physics and Geology, University of Perugia, \\I.N.F.N. Sezione di Perugia,\\
Via Pascoli, I-06123 Perugia, Italy}
\author{  Gordon W. Semenoff}
\affiliation{Department of Physics and Astronomy, University of British Columbia, \\
                     6224 Agricultural Road, Vancouver, British Columbia, Canada V6T 1Z1}
 
\begin{abstract}
We derive a formula for the entanglement entropy of two regions in momentum space that is
generated by the scattering of weakly interacting scalar particles. We discuss an example where
weak interactions entangle momentum scales above and below an infrared cutoff.  
\end{abstract}

\maketitle

 Recently, some insight into the structure of interacting quantum field theories, and in particular conformal field theories \cite{Calabrese:2004eu}
 has been gathered by studying quantum information theoretic issues such as the entanglement entropy of degrees of freedom residing
 in different regions of coordinate space.  One interesting offshoot of this set of ideas is the notion of entanglement entropy in momentum
 space.   From one point of view, there would seem to be some natural questions to address.  The separation of the degrees of freedom  
 in a quantum field theory according to the
 distance scale is a well known ingredient of the Wilson approach to the renormalization group where one might wonder
 whether, even after the construction of effective field theory describing physics at a given distance scale,  subtle quantum effects could still correlate
physics at that scale with physics at other far disparate scales \cite{Balasubramanian:2011wt}.   
On the other hand, interacting quantum field theory is not local in momentum space and the precise meaning of
the degrees of freedom residing in a region of momentum space is a subtle question.   
For example, there exist   computations of the entanglement
of momentum space regions in weakly coupled field theory using perturbation theory and the interaction 
picture  \cite{Balasubramanian:2011wt}\cite{Hsu:2012gk}.  Those compute the entanglement of interaction picture degrees of freedom. The interaction picture fields evolve in time as free fields, 
but the quantum field theory vacuum is a  nontrivial correlated state of those free fields and the correlations include quantum entanglement. 
 However, we could as well consider a different representation of the field theory degrees of freedom, say the Lehman-Symanzik-Zimmerman (LSZ)  in- or out-fields. They are distinguished by the fact that  their states diagonalize the Hamiltonian.  However, because the in- or out-fields  are free fields and their vacuum coincides with the interacting quantum field theory vacuum,  the entanglement entropy  between particle momenta is identical
 to that in free field theory. 
 
 For in- and out-fields, the interactions are encapsulated in the S-matrix of the quantum field theory. 
 An in-state evolves into an out-state, which is a superposition of in-states, and whose individual
coefficients in that superposition are the transition amplitudes, or S-matrix elements. Thus, generally, 
the out-states are entangled states of in-fields and one might naturally ask the question of whether we can
quantify the entanglement that is generated between different regions of momentum space.  
  In this paper, we will develop some formulae for computing this entanglement  in the 
simple example of the self-interacting scalar $\phi$-four theory in four dimensions.  This issue has already been explored in several places
and formulae similar to what we derive are already known  \cite{Seki:2014pca}-\cite{Carney:2016tcs}.  
Our derivation is somewhat different and simpler than previous ones.

We shall consider a pure in-state with $\rho_{\rm in}=|{\rm in}><{\rm in}|$ which evolves to an out-state 
$$
\rho_{\rm out}=|{\rm out}><{\rm out}| = S^\dagger|{\rm in}><{\rm in}|S
$$
where $S$ is the $S$-matrix.  
The  Hilbert space ${\mathcal H}$ of in-states can
be presented as a direct product ${\mathcal H}={\mathcal H}^{\{1\}}\bigotimes{\mathcal H}^{\{2\}}$ where ${\mathcal H}^{\{1\}}$ contains
in-field particles with spatial momenta only in a region $\{1\}\subset R^3$ and ${\mathcal H}^{\{2\}}$ has particles with spatial  momenta in its 
complement $\{2\}=R^3-\{1\}$.  We will produce a reduced density matrix by tracing out all of the in-field states of particles with momenta in
  $\{2\}$
$$
\rho_{\rm red}^{\{1\}}={\rm Tr}_{\{2\}}\rho_{\rm out}
$$
 The reduced density matrix  in an operator in ${\mathcal H}^{\{1\}}$. 
 The entanglement entropy is 
 $$
 S_{\rm ent}=-{\rm Tr}_{\{1\}}\rho_{\rm red}^{\{1\}}\ln\rho_{\rm red}^{\{1\}}
 $$
   For example, with an in-coming two-particle state with momenta $p_1,p_2\in\{1\}$, 
 we shall find that the leading term in the entanglement entropy is 
\begin{align}
S_{\rm ent }=\left[ -\lambda^2 \ln\lambda^2\right] \frac{T}{V}  \frac{1}{\lambda^2}\Sigma_{p_1p_2}+{\mathcal O}(\lambda^2)
\label{entanglement_entropy}
\end{align}
which is  $\left[ -\lambda^2 \ln\lambda^2\right]$, 
times the factor of total time over volume $\frac{T}{V}$, interpreted as particle flux divided by the  relative velocity, 
times the total cross section for scattering at least one particle to a momentum state in $\{2\}$, 
\begin{align}
\Sigma_{p_1p_2}= \frac{\lambda^2 }{256\pi^2} \int_{\{1+2\}}d^3q\int_{\{2\}}d^3k \frac{  \delta^4(p_1+p_2-q-k)}{ \omega(p_1)\omega(p_2)\omega(q)
\omega(k)  }  
\label{cross_section}
\end{align}
which is of order $\lambda^2$. 
The source of this leading perturbative term in the entanglement entropy is the buildup of the eigenvalues of the reduced density
matrix, from those which were zero in the incoming state to transition probabilities to possible out-states which contain 
particles in $\{2\}$. It is interesting that, as in other contexts where entanglement entropy is associated with an area, the entropy here
is proportional to the cross-section. 
We will now present a derivation of equations (\ref{entanglement_entropy}) and
 (\ref{cross_section}). 
After that we will explore an application of this formula.   In any quantum field theory with massless particles, the result of 
a scattering event produces soft particles which are beyond the resolution of detectors and which are not measured.  This leaves
an uncertainty in the observable parts of the process which we could characterize by finding the entanglement of the visible modes
with those which lie below an infrared momentum cutoff.  We will compute this entanglement entropy to the leading
order,  $\lambda^2\ln\lambda^2$ where $\lambda$ is the coupling constant for elastic, short-ranged two-body interactions.

 We will consider   the example of a single real scalar field with a weak repulsive 
two-body interaction. 
The   LSZ reduction formula for the S-matrix is obtained  by setting $J\to 0$ in the expression 
\begin{equation}\label{S}
{\mathcal S}_J=: e^{\left[   \int d^4 x \varphi_{\rm in}(x)\left(-\partial^2+m^2\right)\frac{\delta}{\delta J(x)} \right]}   :~ Z[J]
\end{equation}
 $Z[J]$ is the generating functional for the time-ordered correlation functions of the quantum field theory.
The in-field is   the weak early time limit \footnote{Weak limit means a limit for all matrix elements
of the operator between normalizable states.},
\begin{align}\varphi_{\rm in}(x)={\rm weak}~\lim_{x^0\to-\infty} \varphi(x)
\end{align}
It obeys the
free wave equation and equal time commutation relation
 \begin{align}
& \left(-\partial^2+m^2\right)\varphi_{\rm in}(x)=0\\
& \left[ \varphi(x),\dot\varphi(y)\right]\delta(x^0-y^0)=i\delta^4(x-y)
 \end{align}
 The in-field can  be
decomposed into positive and negative frequency parts, as well as parts with momenta in region $\{1\}$ and region $\{2\}$, 
$$\varphi(x)= \varphi_{\rm in }^{(+)\{1\}}(x)+ \varphi_{\rm in }^{(-)\{1\}}(x)  + \varphi_{\rm in }^{(+)\{2\}}(x)+ \varphi_{\rm in }^{(-)\{2\}}(x) $$
$$ =\int_{\{1\}} \frac{d^3pe^{-i\omega(p)+i\vec p\cdot\vec x}}{\sqrt{2\omega(p)(2\pi)^3}} a^{\{1\}}_p
+\int_{\{1\}}\frac{d^3pe^{i\omega(p)-i\vec p\cdot\vec x}}{\sqrt{2\omega(p)(2\pi)^3}} a^{\{1\}\dagger}_p
$$
$$+\int_{\{2\}} \frac{d^3pe^{-i\omega(p)+i\vec p\cdot\vec x}}{\sqrt{2\omega(p)(2\pi)^3}} a^{\{2\}}_p
+\int_{\{2\}}\frac{d^3pe^{i\omega(p)-i\vec p\cdot\vec x}}{\sqrt{2\omega(p)(2\pi)^3}} a^{\{2\}\dagger}_p
$$  where $\omega(p)=\sqrt{\vec p^2+m^2}$ and the non-vanishing commutator is
$$
\left[ a^{\{i\}}_p, a^{\{j\}\dagger}_q\right]=\delta^{\{i\}\{j\}}\delta(\vec p-\vec q)
$$ 
 We shall assume that  the quantum field theory is renormalized to that  correlation functions of $\varphi(x)$  are finite,   the pole
 of the two-point function of $\varphi$ is at $-p^2=m^2$, the residue of this pole is one and the coupling constant obeys some condition which we shall not need to specify.  
 These are  enforced by adjusting counter-terms in the 
 Lagrangian density
  \begin{align}
&\mathcal{L}=\mathcal{L}_0+\mathcal{L}_I~,~\mathcal{L}_0=-\frac{1}{2}\partial_\mu\varphi\partial^\mu\varphi-\frac{m^2}{2}\varphi^2 \\
&\mathcal{L}_I=-\frac{\lambda}{4!}\varphi^4+{\rm counterterms}
\end{align}
  The generating functional is 
  \begin{align}\label{W1}
Z[J]& = \frac{ \int [D\varphi] e^{i\int  \left[\mathcal{L}+J \varphi \right]}}{\int [D\varphi] e^{i\int  \mathcal{L}}}
=\frac{e^{i\int \mathcal{L}_I\left(\frac{1}{i}\frac{\delta}{\delta J } \right)} e^{-\frac{1}{2}\int J\Delta J} }
{e^{i\int \mathcal{L}_I\left(\frac{1}{i}\frac{\delta}{\delta J }\right)}  e^{- \frac{1}{2}\int J\Delta J} \biggr \rvert_{J=0} }
\end{align}
where
$\Delta(x,y)=<0|{\mathcal T}\varphi_{\rm in}(x)\varphi_{\rm in}(y)|0>
$.
 Using equations (\ref{S})-(\ref{W1}), we can find the $S$ matrix  to second order in $\lambda$, 
\begin{align}\label{SJ}
& S_J=:e^{\int[-\frac{1}{2} J\Delta J+  i  \varphi_{\rm in} J ] }
\left\{1-iT^{(1)}-T^{(2)}+\ldots\right\}:
\end{align}
where (with  $\varphi_{\rm in}^J(z)\equiv  \varphi_{\rm in}(z)+i\int d^4w\Delta(z,w) J(w)$) 
\begin{align}\label{t1}
T^{(1)}&=\dfrac{\lambda}{4!}\int  d^4 z~:\varphi_{\rm in}^J(z)^4 :
 \\
T^{(2)}&=
 \dfrac{\lambda^2}{2(4!)^2} \int  d^4 z_1  d^4 z_2:\varphi_{\rm in}^J(z_1)^4
\varphi_{\rm in}^J(z_2)^4:
  \nonumber \\    & 
  +  \dfrac{8\lambda^2}{(4!)^2}\int  d^4 z_1d^4 z_2\Delta(z_1,z_2):\varphi_{\rm in}^J(z_1)^3\varphi_{\rm in}^J(z_2)^3:
  \nonumber \\    & 
+ \dfrac{3\lambda^2}{2(4!) } \int  d^4 z_1 d^4 z_2\Delta^2(z_1,z_2):\varphi_{\rm in}^J(z_1)^2\varphi_{\rm in}^J(z_2)^2:
 \nonumber \\    & 
+ \dfrac{2\lambda^2}{ 4! } \int  d^4 z_1  d^4 z_2\Delta^3(z_1,z_2):\varphi_{\rm in}^J(z_1)\varphi_{\rm in}^J(z_2)
 :
 \nonumber \\ & +{\rm counterterms}+\ldots
\label{t2}
\end{align}
 where the ellipses stand for contributions of order $\lambda^3$ and higher.  The counterterms  must be tuned to 
cancel the ultraviolet singularities in the 
products of distributions $[\Delta(z_1,z_2)]^2$ and $[\Delta(z_1,z_2)]^3  $ and so that the pole of the two-point function
is the physical mass and the residue is one. When they are properly tuned, the second-last line in (\ref{t2}), which is quadratic in $ \varphi^J_{\rm in}$, vanishes when $J$ is set to zero.
This is the equivalent to stability of the one-particle state. 
We have assumed that tadpoles, that is, terms with $\Delta(x,x)$, have been canceled by counter-terms. 
  
  Given an in-state, $\rho_{\rm in}=|{\rm in}><{\rm in}|$ with  \footnote{In the factor in front of this expression, we have divided by the spatial volume so that the states are unit normalized.  This should more properly be obtained
 as the limit of sharply peaked wave-packets.}
  $$|{\rm in}>=\left( \frac{(2\pi)^3}{V}\right)^{\frac{k+\ell}{2}}
   a_{p_1}^{\{1\}\dagger }\ldots a_{p_k}^{\{1\}\dagger }~ a_{\hat p_1}^{\{2\}\dagger }\ldots a_{\hat p_\ell}^{\{2\}\dagger}  |0>$$ 
we can find the reduced density matrix as
 \begin{align} 
\rho_{\rm red}^{\{1\}} =     \left. | {\mathcal K}_{\hat p_1}  \ldots{\mathcal K}_{\hat p_\ell} |^2
 e^{\mathcal D}~\Pi_{\{1\}} S^\dagger_J \rho_{\rm in}^{\{1\}} S_{\tilde J} \Pi_{\{1\}}  \right|_{J=0=\tilde J}  
 \end{align}
 where  
  \begin{align}
  & \rho_{\rm in}^{\{1\}}=\left( \frac{(2\pi)^3}{V}\right)^{k}
   a_{p_1}^{\{1\}\dagger }\ldots a_{p_k}^{\{1\}\dagger }
    |0><0| a_{p_1}^{\{1\} }\ldots a_{p_k}^{\{1\}  }  \\
 &{\mathcal D} = \int d^4x d^4 y\bar\Delta(x,y){\mathcal P}(x)\tilde {\mathcal P}(y)
\\
&  {\mathcal K}_p =\sqrt{\frac{(2\pi)^3}{V} }  \int d^4x~  \frac{e^{i\omega(p)x^0-i\vec p\cdot\vec x}  }{\sqrt{(2\pi)^32\omega(p)}}\left({\mathcal P(x)}+\tilde{\mathcal P}(x) \right)
 \\
& \bar\Delta(x,y)=<0|\varphi_{\rm in}^{\{2\}}(x)\varphi_{\rm in}^{\{2\}}(y)|0>
\\
&~~~~~~~= \int_{\{2\}}\frac{d^3p  }{(2\pi)^32\omega(p)} e^{-i\omega(p)(x^0-y^0)+i\vec p\cdot (\vec x-\vec y)}
\\
&{\mathcal P}(x)= (-\partial^2+m^2)  \frac{\delta}{\delta J(x)},~\tilde{\mathcal P}(x)= (-\partial^2+m^2)  \frac{\delta}{\delta\tilde J(x)} 
\\ 
&\Pi_{\{1\}}={\mathcal I}_{\{1\}}\otimes |0>^{\{2\}}<0| \end{align}
  $\Pi_{\{1\}} $  projects onto the vacuum of ${\mathcal H}_{\{2\}}$
and leaves excitations with momenta in $\{1\}$ unchanged.

In the simplest case, when $|{\rm in}>$ contains only states in $\{1\}$, 
\begin{align}
&\rho_{\rm red}^{\{1\}}= \Pi_{\{1\}}|{\rm out}><{\rm out}|\Pi_{\{1\} } +\nonumber \\
&+\frac{\lambda^2}{(3!)^2} \int d^4xd^4y\bar\Delta(x,y):\varphi_{\rm in}^3(x):|{\rm in}><{\rm in}| : \varphi_{\rm in}^{\dagger3}(y):  \nonumber\\
&+\frac{\lambda^2}{8} \int d^4xd^4y \bar\Delta^2(x,y):\varphi_{\rm in}^2(x):|{\rm in}><{\rm in}| : \varphi_{\rm in}^{\dagger2}(y):    \nonumber\\
&+\ldots
\label{density}
\end{align}
Terms which would have $\bar\Delta^3(x,y)$ and $\bar\Delta^4(x,y)$ in (\ref{density}) vanish due to kinematical constraints, 
the fact that energy flows only
from $x$ to $y$ in $\bar\Delta(x,y)$.   For the same reason, the second-last line in (\ref{density}) contains states with one fewer particle than 
$|{\rm in}>$ and the last line contains states with two fewer particles.   
For example, for a two-particle state, 
$|\rm in>= {\frac{(2\pi)^3}{V} }a_{p_1}^{\{1\}\dagger}a_{p_2}^{\{1\}\dagger}|0>$ 
\begin{align}
&\rho_{\rm red}^{\{1\}}= \Pi_{\{1\}}|{\rm out}><{\rm out}|\Pi_{\{1\} } +\nonumber \\&
+\int_{\{1\}} d^3q   
\chi_{p_1p_2}^{q }   a^{\{1\}\dagger}_{q}
 |0><0|
 a^{\{1\} }_{ q}
+\chi_{p_1p_2}     |0><0| +\ldots
\label{density1}
\end{align}
which has eigenvalues
\begin{align}
&\chi_{p_1p_2}^{q }=\frac{ \lambda^2  }{256\pi^2}\frac{(2\pi)^3}{V}\frac{T}{V}
\int_{\{2\}} \frac{ d^3k\delta^4(p_1+p_2-q-k)}{ \omega(p_1)\omega(p_2)\omega(q)
\omega(k)  }  
 \\
&\chi_{p_1p_2} =\frac{ \lambda^2  }{256\pi^2}\frac{T}{V}\int_{\{2\}} \frac{ d^3qd^3k\delta^4(p_1+p_2-q-k)}{ \omega(p_1)\omega(p_2)\omega(q)
\omega(k)  }    \\
&1-\frac{ \lambda^2  }{256\pi^2}\frac{T}{V}\int_{\{1\}+\{2\}}d^3q\int_{\{2\}}d^3k \frac{  \delta^4(p_1+p_2-q-k)}{ \omega(p_1)\omega(p_2)\omega(q)
\omega(k)  }  
\end{align}
where $V$ is the volume and $T$ the total time. 
The leading term in the entanglement entropy is just
the expression given in equation (\ref{entanglement_entropy}) and (\ref{cross_section}).

 Now, let us consider an application of equation (\ref{entanglement_entropy}) where the particles are massless and 
 $\{1\}=\{\vec p~:~|\vec p|>\Lambda\}$ and   $\{2\}=\{ \vec p~ :~ |\vec p|\leq \Lambda\}$, with $\Lambda$   an infrared cutoff.  The idea is that the cutoff is the resolving
   power of particle detectors, which is always finite, since, for example,  
   they should have finite size and the excitations with wave-lengths
   larger than their size would escape detection. In any scattering event, some soft particles, 
   with momenta smaller than $\Lambda$ could be produced and this is tantamount to information loss.   
 The entropy of the entanglement of those degrees of freedom which remain above the cutoff to those below
   the cutoff is a quantitative measure of information loss. The for massless particles, the elementary integral 
   in (\ref{entanglement_entropy}) gives 

  \begin{align}
S_{\rm ent}&= \frac{ \lambda^2\ln\lambda^2  }{128\pi}\frac{T}{V}
\frac{ \Lambda -\frac{1}{2}\left[ |\vec p_1|+|\vec p_2|- |\vec p_1 +\vec p_2|\right] }{|\vec p_1||\vec p_2||\vec p_1+\vec p_2|}\\
&{\rm when }~~
 \Lambda \geq \frac{1}{2}\left[ |\vec p_1|+|\vec p_2|- |\vec p_1 +\vec p_2|\right]  ~~ |\vec p_1|,|\vec p_2|\geq\Lambda \nonumber
\end{align}
Since the cutoff on spatial momentum is not covariant,  the amount of entropy generated is frame dependent. What is more, it is
constrained by kinematics as the production of particles below the cutoff must be allowed by the conservation of energy and momentum. 
This depends
sensitively on the angle, $\chi$  between the incident $\vec p_1$ and $\vec p_2$ as well as the magnitudes $p_1$ and $p_2$.  
The entropy will be non-zero if 
$$\sin^2\chi/2   \leq  \Lambda( \frac{1}{ p_1} + \frac{1}{p_2} )- \frac{ \Lambda^2 }{p_1p_2}$$
The right-hand-side of this equation is positive for all values of $p_1,p_2>\Lambda$.  It is equal to one, which allows any angle, when
$p_1=p_2=\Lambda$ and it decreases from one as either of $p_1$ or $p_2$ are increased into region $\{1\}$. This tells us that $\chi=\pi$
radians is not allowed for any $p_1,p_2>\Lambda$.  This excludes entropy generation in the center of mass frame.  
  
  It would be interesting to generalize our computations to higher orders in perturbation theory.  For example, 
  this would be needed to explore the scenario where
  the coupling constant runs.  For example, the infrared freedom of phi-four theory with an ultraviolet cutoff could make the infrared
  sector solvable. It is also interesting to ask if the idea of entanglement with infrared degrees of freedom is relevant to the infrared problem
  in massless gauge field theories.   
  
 \vskip .5cm
 The authors thank NSERC of Canada and the INFN of Italy for financial support.  G.G. acknowledges the hospitality of the
 University of British Columbia, where this work initiated and G.W.S. the hospitality of the INFN, Sezione di Perugia and the University
 of Perugia, where
 it was completed. We thank Dan Carney and Laurent Charette for informative discussions. 
   


\end{document}